\documentclass[10pt, conference]{IEEEtran}
\IEEEoverridecommandlockouts
\usepackage{cite}
\usepackage{amsmath,amssymb,amsfonts}
\usepackage{algorithmic}
\usepackage{graphicx}
\usepackage{textcomp}
\usepackage{xcolor}
\usepackage{booktabs}
\usepackage{multirow}
\usepackage{balance}
\usepackage{hyperref}
\usepackage{listings}
\hypersetup{
  colorlinks,
  linkcolor=red, 
  urlcolor=blue,
  citecolor=black,
  bookmarksnumbered,
  unicode
}
\def\BibTeX{{\rm B\kern-.05em{\sc i\kern-.025em b}\kern-.08em
    T\kern-.1667em\lower.7ex\hbox{E}\kern-.125emX}}
\begin{document}


\title{Evaluating the Generalizability of LLMs in Automated Program Repair}

\author{\IEEEauthorblockN{Fengjie Li\IEEEauthorrefmark{1}, Jiajun Jiang\IEEEauthorrefmark{1}\IEEEauthorrefmark{2}\thanks{\IEEEauthorrefmark{2}Jiajun Jiang is the corresponding author for this work.}, Jiajun Sun\IEEEauthorrefmark{1}, Hongyu Zhang\IEEEauthorrefmark{3}}
\IEEEauthorblockA{\IEEEauthorrefmark{1}College of Intelligence and and Computing, Tianjin University, Tianjin, China}
\IEEEauthorblockA{\IEEEauthorrefmark{3}School of Big Data and Software Engineering, Chongqing University, Chongqing, China}
\IEEEauthorblockA{\IEEEauthorrefmark{1}\{fengjie, jiangjiajun, sjjtianjin\}@tju.edu.cn, \IEEEauthorrefmark{3}hyzhang@cqu.edu.cn}
}


\maketitle

\newcommand{\DATASET}[1]{\textsc{Defects4J-Trans}}

\newcommand{\jiajun}[1]{{\color{red}[Jiajun: #1]}}

\newcommand{\fengjie}[1]{{\color{brown}[Fengjie: #1]}}

\newcommand{\hy}[1]{{\color{red}[HY: #1]}}

\newcommand{\add}[1]{{\color{blue}#1}}
\newcommand{\del}[1]{{\color{red}\sout{#1}}}

\renewcommand{\add}[1]{#1}

\lstdefinestyle{java}{ 
	language=java,
	basicstyle=\scriptsize\ttfamily\bfseries\color[rgb]{0.0, 0.5, 0.69}, 
	breakatwhitespace=false, 
	breaklines=true, 
	captionpos=b, 
	commentstyle=\color[rgb]{0.0, 0.5, 0.69},
	deletekeywords={}, 
	escapeinside={<@}{@>},
	firstnumber=1, 
	frame=shadowbox, 
	frameround=none, 
	keywordstyle={[1]\color{blue!90!black}},
	keywordstyle={[3]\color{red!80!orange}},
	morekeywords={String,int}, 
	numbers=none, 
	numbersep=-8pt, 
	numberstyle=\tiny\color[rgb]{0.1,0.1,0.1}, 
	rulecolor=\color{black}, 
	showstringspaces=false, 
	showtabs=false, 
	stepnumber=1, 
	stringstyle=\color[rgb]{0.58,0,0.82},
	tabsize=2, 
	backgroundcolor=\color{white}
}

\newcommand{\hlc}[2]{{\setlength\fboxsep{0pt}\hspace{-3pt}\colorbox{#1} 
		{\begin{minipage}{\dimexpr\columnwidth-1\fboxsep+0pt\relax}
				\textcolor{black}{\textbf{\strut\hspace{3pt}#2}}
			\end{minipage}}}}

\begin{abstract}
LLM-based automated program repair methods have attracted significant attention for their state-of-the-art performance. However, they were primarily evaluated on a few well-known datasets like Defects4J, raising questions about their effectiveness on new datasets. In this study, we evaluate 11 top-performing LLMs on \DATASET{}, a new dataset derived from transforming Defects4J while maintaining the original semantics. Results from experiments on both Defects4J and \DATASET{} show that all studied LLMs have limited generalizability in APR tasks, with the average number of correct and plausible patches decreasing by 49.48\% and 42.90\%, respectively, on \DATASET{}. Further investigation into incorporating additional repair-relevant information in repair prompts reveals that, although this information significantly enhances the LLMs' capabilities (increasing the number of correct and plausible patches by up to 136.67\% and 121.82\%, respectively), performance still falls short of their original 
results. This indicates that prompt engineering alone is insufficient to substantially enhance LLMs' repair capabilities. Based on our study, we also offer several recommendations for future research.

\end{abstract}


\section{Introduction}
\label{sec:intro}

With the rapid growth of the scale and complexity of modern software systems, the number and intricacy of software bugs have also increased, resulting in significant financial 
losses for organizations and end-users. Fixing these bugs requires substantial consumption of time and effort from developers. As a result, Automated Program Repair~(APR), which focuses on automatically fixing software bugs~\cite{jiang2018shaping,liu2019tbar,zhu2021syntax, jiang2023knod, zhu2023tare,jiang2019inferring}, has attracted considerable attention from academia and industry.


Recently, Large Language Models (LLMs) have demonstrated impressive performance across various software engineering (SE) tasks leading to the emergence of an increasing number of LLM-based APR methods~\cite{Xia2022LessTM, xia2023plastic, zhang2023gamma, xiang2024far, silva2023repairllama, xia2024automated, yin2024thinkrepair}. 
Several existing works~\cite{Xia2022LessTM,xia2023automated,jiang2023impact} have shown that LLMs, even without additional repair-relevant information, can outperform previous learning-based APR methods just using a few-shot prompting. Researchers have explored various approaches to enhance LLM-based APR by providing more repair-relevant information, such as fault localization~\cite{jiang2023impact}, bug reports~\cite{xiang2024far} and trigger test information~\cite{xia2024automated}. Techniques such as Chain-of-Thought (CoT) prompting~\cite{wei2022chain}, multi-turn dialogues\cite{yi2024survey}, and multi-agent systems~\cite{xi2023rise} have been employed to better inform LLMs and improve their capabilities.

Although these LLM-based APR methods have achieved remarkable results, they have primarily only been evaluated on well-known datasets such as Defects4J~\cite{just2014defects4j} and QuixBugs~\cite{lin2017quixbugs}, which were proposed years ago. Recent works~\cite{zhang2023critical, lopez2024inter, sallou2024breaking,  xu2024benchmarking} have highlighted the significant risk of memorization in LLMs when evaluated on these datasets, leading to our \textit{\textbf{first research question~(RQ1): How is the generalizability of LLMs?}} Specifically, can LLMs achieve the same impressive performance on another fresh new dataset? \add{To answer this question, we create the \DATASET{} dataset by applying program transformations to Defects4J dataset, ensuring emantic equivalence while altering code content.}
Then, we assess the generalizability of LLMs by comparing their repair results on both Defects4J and \DATASET{}.


\noindent \textbf{\underline{Results:}} The results show that the performance of LLMs significantly declined on \DATASET{}, with the number of correct and plausible patches decreasing by an average of \textbf{49.48\%} and \textbf{42.90\%}, respectively. This suggests unsatisfactory generalizability and underscores the need for more comprehensive evaluations of LLM-based APR methods.

This poor generalizability raises our \textit{\textbf{second research question~(RQ2): Can repair-relevant information enhance the generalizability of LLMs?}} To answer this question, we incorporate three types of repair-relevant information used in previous LLM-based APR methods~\cite{xia2023plastic, xia2024automated, yin2024thinkrepair, xiang2024far}, to explore their potential to enhance the generalizability of LLMs.

\noindent \textbf{\underline{Results:}} 
The results indicate that incorporating this information increased the number of correct and plausible patches by up to \textbf{136.67\%} and \textbf{121.82\%}, respectively. However, most LLMs still underperformed compared to their original results on Defects4J, highlighting the need to develop more effective LLM-based APR methods.


\noindent \textbf{Contributions.} To sum up, the contributions of this study are:

\noindent$\bullet$ We conducted the first extensive experiments to investigate the generalizability of 11 top-performing LLMs, 
using 4 types of prompts to explore the impact of different information.
Our findings reveal that the selected LLMs exhibit unsatisfactory generalizability in APR tasks.

\noindent$\bullet$ We propose several future directions to enhance the generalizability of LLM-based methods. Our findings aim to guide the SE community in effectively utilizing LLMs for APR.

\noindent$\bullet$ We have open-sourced all code, data and results involved in our work~\cite{homepage} to promote replication and future research.

\section{Related Work}
\label{sec:background}
We have not identified any studies dedicated to validating the generalizability of LLMs in the APR task in the literature, although some works aimed at addressing the risks associated with LLMs in SE tasks. For example, studies~\cite{zhang2023critical, lopez2024inter, sallou2024breaking, xu2024benchmarking} have pointed out several potential issues (e.g., memorization and reproducibility) when using LLMs for SE tasks, but they lack thorough experimental validation. Additionally, new defect datasets have been proposed to address the memorization issues in LLMs, such as ConDefects~\cite{10.1145/3663529.3663815}, which collected recent bugs from the online competition platform, and HumanEval-Java~\cite{jiang2023impact}, a mutation of the original HumanEval~\cite{chen2021evaluating} dataset. However, these datasets consist mainly of simple algorithmic programs, where LLMs tend to perform well even when directly generating the corresponding functional code, limiting their ability to fully reflect the LLMs’ coding capabilities. In contrast, \DATASET{} used in this study retains more characteristics of real-world projects, such as project-specific APIs, while effectively mitigating memorization issues in LLMs. Other works~\cite{li2024evocodebench,xia2024top} have also introduced evolving datasets for code generation tasks, which are different from our study.

\section{study design}
\label{sec:design}

\begin{figure}[tbp]
    \centering
    \includegraphics[width=\columnwidth]{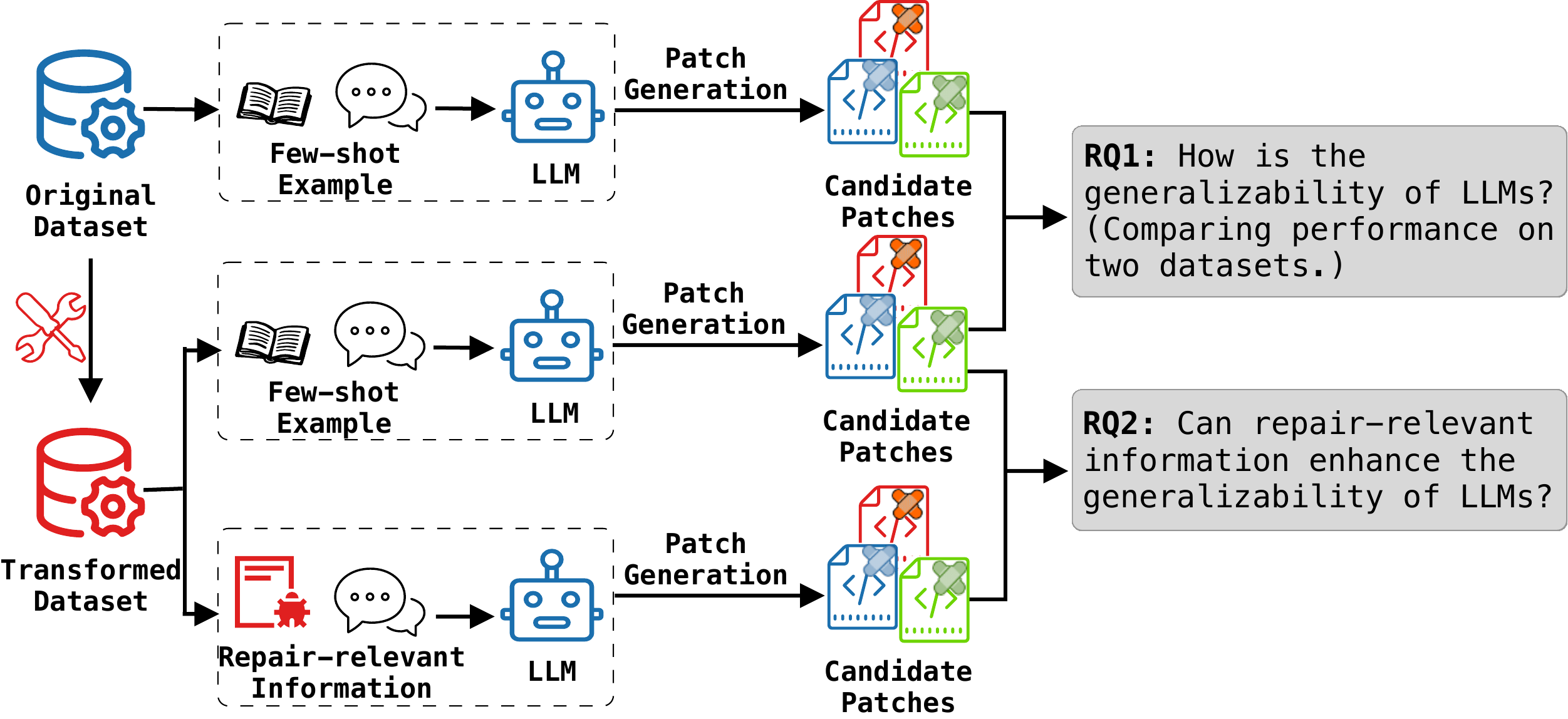}
    \caption{Overview of study design}
    \label{fig:overview}
\end{figure}

In this section, we detail the LLMs selected for evaluation, the construction of \DATASET{}, the utilized prompts, and the overall settings of the LLMs employed in this study.
Figure~\ref{fig:overview} shows the overview of our study. First, we apply our code transformation tool to the Defects4J dataset to generate \DATASET{}, which maintains the original fault semantics but contains different code content. Then, we apply the LLMs on both Defects4J and \DATASET{}, comparing their performance differences to assess their generalizability. Finally, we investigate the impact of three types of repair-related information on the generalizability of LLMs.

\subsection{Studied LLMs}
\label{sec:design_sub_llms}

Our selection of LLMs is based on the EvalPlus leaderboard~\cite{liu2024your, EvalPlusLeaderBoard}, which offers more rigorous tests for evaluating coding capabilities of LLMs. We selected LLMs that ranked in the top 20 on that leaderboard and capable of performing inference locally. Due to resource constraints, we excluded LLMs with parameter size exceeding 34B and those without open-source access. In total, we chose 11 different LLMs for our evaluation.
\add{These remaining LLMs still exhibit competitive coding capabilities. For instance, CodeQwen1.5 achieved a score of 73.8, securing the 4th position on the leaderboard.} Table~\ref{tab:llm_overview} provides details, including LLM name, number of parameters, and average pass@1 score on the  
EvalPlus benchmark.
As shown, the parameter sizes of LLMs ranging from 6.7B to 34B\add{, with scores varying from 66.5 to 73.8. Particularly, GPT-4-Turbo achieves a score of 77.5}.

\begin{table}[tb]
    \centering
    \caption{Summary of studied LLMs.}
    \begin{tabular}{lcc}
    \toprule
         LLM & \#Parameters $\uparrow$ & \#Average Score \\
    \midrule
         WaveCoder-Ultra~\cite{yu2024wavecoder} & 6.7B  & 66.5 \\
         DeepSeek-Coder-Instruct~\cite{guo2024deepseek} & 6.7B  & 68.4 \\
         OpenCodeInterpreter-DS~\cite{zheng2024opencodeinterpreter} & 6.7B  & 69.2 \\
         Magicoder-S-DS~\cite{wei2024magicoder} & 6.7B  & 70.2 \\
         Artigenz-Coder-DS~\cite{Artigenz-Coder} & 6.7B  & 71.1 \\
         DeepSeek-Coder-Instruct-v1.5~\cite{guo2024deepseek} & 7B  & 66.8 \\
         CodeQwen1.5~\cite{qwen} & 7B & 73.8 \\
         StartChat2-v0.1~\cite{Tunstall_The_Alignment_Handbook} & 15B  & 67.9 \\
         OpenCodeInterpreter-DS~\cite{zheng2024opencodeinterpreter} & 33B  & 71.2\\
         DeepSeek-Coder-Instruct~\cite{guo2024deepseek} & 33B  & 72.5\\
         SpeechLess-CodeLlama-v2.0~\cite{SpeechLess-CodeLlama-v2.0} & 34B  & 66.7 \\
    \bottomrule
    \end{tabular}
    \label{tab:llm_overview}
\end{table}

\subsection{Construction of \DATASET{}}
\label{sec:design_sub_benchmarks}

To ensure the semantic equivalence, we primarily performed equivalent transformations related to control flow statements in the code. Following existing code transformation works~\cite{cheers2019spplagiarise, tian2023code}, we designed five transformation operators and applied them by parsing and modifying abstract syntax trees~(AST) using the Java Development Toolkit~\cite{JDTAST}. Due to space limits, we briefly introduce each transformation operator’s functionality here. For detailed implementation, please refer to our homepage~\cite{homepage}:


\begin{itemize}
    \item \textbf{T$_{1}$-Variable Renaming:} Replaces all variable names in the original code with the new names generated by StarCoder2-3B to maintain the naturalness of code. 
    
    \item \textbf{T$_{2}$-Loop Transformation:} Transforms between \texttt{For-Loop} and \texttt{While-Loop} structures  equivalently.
    
    \item \textbf{T$_{3}$-Switch Transformation:} Replaces \texttt{Switch Statement} with a series of \texttt{If Statement} by adding \texttt{break flag} and \texttt{fall-through flag} on demand to ensure consistent executing logic. 
    
    \item \textbf{T$_{4}$-Dead Code Injection:} Inserts statements that will never be executed, such as \texttt{if (false) \{...\}}.  To save tokens, we applied this operator once under each \texttt{Block Statement}, and injected at most three instances of dead code within a single function.
    
    \item \textbf{T$_{5}$-Boolean Transformation:} Double negating boolean predicates in \texttt{If Statement}s.
    For example, \texttt{if (!(!(condition))) \{...\}}.
\end{itemize}

We sequentially apply the above five operators to all 438 single-function bugs in Defects4J (existing LLM-based APR methods mainly focusing on repairing bugs within a single function~\cite{xia2023automated, xiang2024far, xia2024automated}) to generate \DATASET{}. Semantic consistency between \DATASET{} and the original Defects4J was ensured through manual inspection and by running Defects4J's compilation and testing scripts.


\begin{lstlisting}[style=java, numbers=none, label=prompt,caption=The input prompt used in this study.]
<@\hlc{gray!25}{// Provide a fix for the buggy function} @>
{Buggy code and fixed code exmaples} or
{Repair-relevant Information}
<@\hlc{gray!25}{// Provide a fix for the buggy function} @>
<@\hlc{gray!25}{// Buggy Function} @>
<@\textcolor{red}{\{Buggy code want to fix\}} @>
<@\hlc{gray!25}{//Fixed Function} @>
\end{lstlisting}


\subsection{Prompt Engineering}
\label{sec:design_sub_prompt}

Listing~\ref{prompt} shows the prompts used in this study. Following prior works~\cite{xia2023automated, xiang2024far}, we provide the entire buggy function as the input to the LLMs, which then generate the complete fixed function. This approach reflects a practical scenario and effectively assesses the coding capabilities of the LLMs. Specifically, 
the basic prompt template employs ``\texttt{//Provide a fix for the buggy function}'' to indicate the APR task and uses ``\texttt{//Buggy Function}'' and ``\texttt{//Fixed Function}'' to help the LLM identify buggy and fixed code. Additionally, for evaluating the impact of additional repair-relevant information, we designed three extended prompts, detailed as follows.

\textbf{1) Two-shot:} Using the aforementioned template, we provide two pairs of buggy and fixed code examples. One is a manually constructed toy example to help LLMs understand the APR task, and the other is the repair example with the shortest context from the same project, offering insight into the coding style. This serves as the default prompt in RQ1. 

\textbf{2) Two-shot$_{fl}$:} This prompt further
incorporates perfect line-level fault localization information into the aforementioned Two-shot prompt by manually annotating the buggy lines in the functions with \texttt{/*bug is here*/}.


\textbf{3) Bug Report:} This prompt replaces the two-shot examples with bug report information, and marks the report title and content with ``\texttt{//Bug Report Title}'' and ``\texttt{//Bug Report Content}''. 
In other words, we only provide the faulty function and the associated bug report to fit the input length limit of LLMs without presenting repair examples.

\textbf{4) Trigger Test:} Similarly, this prompt replace the repair examples with the trigger test and the corresponding error message, which are marked with
``\texttt{//Trigger Test}'' and ``\texttt{//Error Message}'', respectively.

\subsection{General Settings}
\label{sec:design_sub_settings}

We used the HuggingFace Library~\cite{HuggingFace} to load LLM weights and perform inference. Following Xia et al.~\cite{xia2023automated}, we set top-p to 0.95 and the temperature to 0.8. Each LLM was invoked 200 times with the respective prompt for each bug. A patch is considered \textit{plausible} if it passes the test cases and \textit{correct} if it is semantically equivalent to the developer’s patch.

Our experiments were conducted on a local machine equipped with dual Intel Xeon Golden 6348 CPUs, 512GB RAM, and eight A800 GPUs, running Ubuntu 20.04.6LTS.

\subsection{Research Questions}
\label{sec:design_sub_rq}

In this work, we study the following two questions:

\begin{itemize}
\setlength{\itemindent}{0pt} 
    \setlength{\leftskip}{0pt} 
    \item \textbf{RQ1:How is the generalizability of LLMs?} We 
    evaluated the generalizability of 11 top-performing LLMs in program repair by comparing their repair capabilities on Defects4J and \DATASET{} accordingly.
    
    \item \textbf{RQ2:Can repair-relevant information enhance the generalizability of LLMs?} Based on RQ1, we selected 4 LLMs that exhibited the weakest generalizability in this RQ to optimize time efficiency. We empirically analyzed whether incorporating repair-relevant information could enhance their repair performance.
\end{itemize}

\section{Preliminary evaluation}
\label{sec:evaluation}

\subsection{RQ1: Generalizability}
\label{sec:evaluation_sub_generalizability} 


\begin{table}[tbp]
\centering
\caption{Repair performance of Different LLMs on Defects4J (D4J) and \DATASET{} (D4J-T)}
\label{tab:table_1_decrease}
\resizebox{0.98\columnwidth}{!}{
\begin{tabular}{l|c|ccc|ccc}
\toprule
\multirow{2}{*}{LLM} & \multirow{2}{*}{\#Parameters} & \multicolumn{3}{c|}{\#Correct Fixes}  & \multicolumn{3}{c}{\#Plausible Fixes}\\ 
  & & \#D4J & \#D4J-T & Dec.(\%) & \#D4J & \#D4J-T & Dec.(\%)\\
\midrule
WaveCoder-Ultra&6.7B&65&34&47.69$\downarrow$&105&57&45.71$\downarrow$\\
DeepSeek-Coder-Instruct&6.7B&63&36&42.86$\downarrow$&114&73&35.96$\downarrow$\\
OpenCodeInterpreter-DS&6.7B&56&27&51.79$\downarrow$&91&51&43.96$\downarrow$\\
Magicoder-S-DS&6.7B&77&38&50.65$\downarrow$&120&67&44.17$\downarrow$\\
Artigenz-Coder-DS&6.7B&75&43&42.67$\downarrow$&111&71&36.04$\downarrow$\\
CodeQwen1.5&7B&74&37&50.00$\downarrow$&125&72&42.40$\downarrow$\\
DeepSeek-Coder-Instruct-v1.5&7B&71&42&40.85$\downarrow$&136&89&34.56$\downarrow$\\
StarChat2-v0.1&15B&92&30&\textbf{67.39}$\downarrow$&152&55&\textbf{63.82}$\downarrow$\\
OpenCodeInterpreter-DS&33B&87&37&57.47$\downarrow$&119&61&48.74$\downarrow$\\
DeepSeek-Coder-Instruct&33B&94&45&52.13$\downarrow$&133&76&42.86$\downarrow$\\
Speechless-CodeLlama-v2.0&34B&107&66&38.32$\downarrow$&160&108&32.50$\downarrow$\\
\midrule
\multicolumn{2}{c|}{Average}&78.27&39.55&49.48$\downarrow$&124.18&70.91&42.90$\downarrow$\\
\bottomrule
\end{tabular}
}
\end{table}

\begin{table*}[tb]
\centering
\caption{Repair performance of LLMs on \DATASET{} using different prompts}
\label{tab:table_2_prompt}
\resizebox{\textwidth}{!}{
\begin{tabular}{l|cc|ll|ll|ll}
\toprule
\multirow{2}{*}{LLM} & \multicolumn{2}{c|}{Two-Shot}  & \multicolumn{2}{c|}{Two-Shot$_{fl}$} & \multicolumn{2}{c|}{Trigger Test} & \multicolumn{2}{c}{Bug Report}\\
& \#Correct Fixes&\#Plausible Fixes& \#Correct Fixes&\#Plausible Fixes&\#Correct Fixes&\#Plausible Fixes& \#Correct Fixes&\#Plausible Fixes\\
\midrule

WaveCoder-Ultra-6.7B&34&57&44(+29.41\%$\uparrow$)&64(+12.28\%$\uparrow$)&59(+73.53\%$\uparrow$)&94(+64.91\%$\uparrow$)&72(+111.76\%$\uparrow$)&111(+94.74\%$\uparrow$)\\
StarChat2-v0.1-15B&30&55&53(+76.67\%$\uparrow$)&76(+38.18\%$\uparrow$)&71(\textbf{+136.67\%$\uparrow$})&114(+107.27\%$\uparrow$)&69(+130.0\%$\uparrow$)&122(\textbf{+121.82\%$\uparrow$})\\
OpenCodeInterpreter-DS-6.7B&27&51&48(+77.78\%$\uparrow$)&75(+47.06\%$\uparrow$)&43(+59.26\%$\uparrow$)&72(+41.18\%$\uparrow$)&62(+129.63\%$\uparrow$)&100(+96.08\%$\uparrow$)\\
OpenCodeInterpreter-DS-33B&37&61&50(+35.14\%$\uparrow$)&67(+9.84\%$\uparrow$)&59(+59.46\%$\uparrow$)&96(+57.38\%$\uparrow$)&68(+83.78\%$\uparrow$)&110(+80.33\%$\uparrow$)\\
\midrule
\textbf{Average} & 32.00 & 56.00 & 48.75(+52.34\%$\uparrow$) & 70.5(+25.89\%$\uparrow$) & 58.0(+81.25\%$\uparrow$) & 94.0(+67.86\%$\uparrow$) & 67.75(+111.72\%$\uparrow$) & 110.75(+97.77\%$\uparrow$) \\

\bottomrule
\end{tabular}
}
\end{table*}

Table~\ref{tab:table_1_decrease} shows the number of correct and plausible patches generated by 11 LLMs using the Two-shot prompt, as described in Section~\ref{sec:design_sub_prompt}, on both Defects4J and \DATASET{}. We observe a decline in both correct and plausible repairs across all selected LLMs. Notably, the StarChat2~\cite{Tunstall_The_Alignment_Handbook} experienced the largest drop, with correct and plausible repairs decreasing by 67.39\% and 63.82\%, respectively. The smallest decline was seen in SpeechLess-CodeLlama-v2.0~\cite{SpeechLess-CodeLlama-v2.0}, with decreases of 38.32\% and 32.50\%. On average, the number of correct and plausible patches generated by the LLMs decreased by 49.48\% and 42.90\%, respectively. This results indicate that LLMs may still struggle to correctly understand the semantics of faulty programs as their repair capability highly depends on the form of code. 
\add{For instance, we observed that many LLM-generated patches tend to repeatedly modify the injected dead code or transformed boolean predicates, rather than addressing the actual errors. This also motivates the experiments in RQ2.}

When comparing the 6.7B and 33B parameter versions of DeepSeek-Coder-Instruct and OpenCodeInterpreter-DS, we observe a significant scaling effect. LLMs with larger parameter size generate more correct and plausible patches. Interestingly, the larger parameter size appears to correlate with weaker generalization capabilities. The two 33B models experienced decline rates of 52.13\% and 57.47\%, while the decline rates for the 6.7B models were 42.86\% and 51.79\%. We believe this is due to the larger models having a higher degree of overfitting to the original dataset.

\subsection{RQ2: Impacts of Repair relevant Information}
\label{sec:evaluation_sub_impact}

Table~\ref{tab:table_2_prompt} shows the repair results of the four selected LLMs, which exhibited the most significant decline in RQ1~(\ref{sec:evaluation_sub_generalizability}), using different prompts. Overall, with the incorporation of additional repair-relevant information, all LLMs demonstrate improved repair capabilities, with 29.41\% to 136.67\% improvements on the number of correct patches and 9.84\% to 121.82\% improvements on the number of plausible patches.

Among the three types of repair-relevant information added, the bug report information yielded the most significant improvement for these four LLMs, with average increases of 111.72\% in correct patches and 97.77\% in plausible patches. We attribute this enhancement to the high quality of the bug reports in the Defects4J dataset, which involved substantial human effort in analyzing and providing information relevant to bug identification and repair, allowing the LLMs to better understand how to fix the buggy code. In contrast, fault localization information provided the least improvement for the four LLMs, with average increases of 52.34\% in correct patches and 25.89\% in plausible patches. This aligns with previous findings~\cite{jiang2023impact}, suggesting that LLMs may struggle to effectively interpret fault localization information in this format. A possible reason is the limited availability of training data that include fault localization identifiers within the code.

Although incorporating repair-relevant information can improve LLMs' repair capabilities,
only OpenCodeInterpreter-6.7B and WaveCoder-Ultra-6.7B showed slight improvements with bug reports across 12 experimental setups (4 LLMs $\times$ 3 prompts), compared to their performance
on the original Defects4J dataset. The other 10 experimental setups still performed below their results on the original dataset. The findings suggest that enhancing the generalizability of LLM-based APR methods still requires significant progress.




\section{Threats to Validity}
\label{sec:threats}
 Manually reviewing all plausible patches to identify correct patches that are semantically consistent with the developer patches is an internal threat to the validity of our work. Following common APR practice, we perform a careful analysis of each plausible patch and have published our full set of correct and plausible patches at our homepage~\cite{homepage}.


\section{future plans}
\label{sec:future}

\textbf{Evaluate on a broader range of LLM-based APR methods and datasets.} The number of LLM-based methods selected in this study is limited, and the investigation was conducted solely on Java programming language. In the future, we aim to evaluate more LLM-based methods on more datasets across different programming languages.

\add{\textbf{Explore better methods to enhance LLMs’ understanding of program semantics.} This study reveals that LLMs continue to face challenges in accurately grasping program semantics. Most LLM-based APR methods focus primarily on designing different ways to utilize LLMs. A promising direction lies in integrating traditional APR techniques, such as heuristic and constraint-solving based methods, with LLMs. For example, program analysis and verification could be applied to validate and refine LLM-generated patches, thereby improving their semantic understanding and repair capabilities.}

\textbf{Enhance the generalizability of LLMs.} This study indicates that simple code transformation can significantly degrade the performance of LLMs. 
Therefore, we plan to explore a code normalization method to unify code format before processing by LLMs. 
Specifically, we may either pre-define a set of code normalization rules or use a fine-tuned LLM to ensure consistent representation of code, such as naming conventions and code structures, with the same semantics. This approach potentially allows us to filter out non-essential code features while preserving the key code semantics. Moreover, the unified representation can be used to fine-tune LLMs and enhance the application of learned knowledge from historical data.
Finally, we urge developers of LLMs and LLM-based methods to share their datasets and engage in discussions about potential risks.

\section{conclusion}
\label{sec:conclusion}
In this study, we explore the generalizability of LLMs in the APR task and examine how various types of repair-relevant information affect the LLMs’ bug-fixing capabilities.
Our findings indicate that the studied 11 top-performing LLMs show limited genralizability. While incorporating repair-relevant information helps improve repair performance, challenges remain. Based on our study, we propose several promising directions
for future research 
and have open-sourced all our experimental data to facilitate replication and further investigation~\cite{homepage}.


\section{Acknowledgment}

We thank the anonymous reviewers for their constructive suggestions to help improve the quality of this paper. 
This work was supported by the National Natural Science Foundation of China under Grant Nos. 62202324.

\balance
\bibliographystyle{IEEEtran}
\bibliography{IEEEabrv,reference}

\begin{thebibliography}{10}
\providecommand{\url}[1]{#1}
\csname url@samestyle\endcsname
\providecommand{\newblock}{\relax}
\providecommand{\bibinfo}[2]{#2}
\providecommand{\BIBentrySTDinterwordspacing}{\spaceskip=0pt\relax}
\providecommand{\BIBentryALTinterwordstretchfactor}{4}
\providecommand{\BIBentryALTinterwordspacing}{\spaceskip=\fontdimen2\font plus
\BIBentryALTinterwordstretchfactor\fontdimen3\font minus \fontdimen4\font\relax}
\providecommand{\BIBforeignlanguage}[2]{{%
\expandafter\ifx\csname l@#1\endcsname\relax
\typeout{** WARNING: IEEEtran.bst: No hyphenation pattern has been}%
\typeout{** loaded for the language `#1'. Using the pattern for}%
\typeout{** the default language instead.}%
\else
\language=\csname l@#1\endcsname
\fi
#2}}
\providecommand{\BIBdecl}{\relax}
\BIBdecl

\bibitem{jiang2018shaping}
J.~Jiang, Y.~Xiong, H.~Zhang, Q.~Gao, and X.~Chen, ``Shaping program repair space with existing patches and similar code,'' in \emph{Proceedings of the 27th ACM SIGSOFT international symposium on software testing and analysis}, 2018, pp. 298--309.

\bibitem{liu2019tbar}
K.~Liu, A.~Koyuncu, D.~Kim, and T.~F. Bissyand{\'e}, ``Tbar: Revisiting template-based automated program repair,'' in \emph{Proceedings of the 28th ACM SIGSOFT international symposium on software testing and analysis}, 2019, pp. 31--42.

\bibitem{zhu2021syntax}
Q.~Zhu, Z.~Sun, Y.-a. Xiao, W.~Zhang, K.~Yuan, Y.~Xiong, and L.~Zhang, ``A syntax-guided edit decoder for neural program repair,'' in \emph{Proceedings of the 29th ACM joint meeting on European software engineering conference and symposium on the foundations of software engineering}, 2021, pp. 341--353.

\bibitem{jiang2023knod}
N.~Jiang, T.~Lutellier, Y.~Lou, L.~Tan, D.~Goldwasser, and X.~Zhang, ``Knod: Domain knowledge distilled tree decoder for automated program repair,'' in \emph{2023 IEEE/ACM 45th International Conference on Software Engineering (ICSE)}.\hskip 1em plus 0.5em minus 0.4em\relax IEEE, 2023, pp. 1251--1263.

\bibitem{zhu2023tare}
Q.~Zhu, Z.~Sun, W.~Zhang, Y.~Xiong, and L.~Zhang, ``Tare: Type-aware neural program repair,'' in \emph{2023 IEEE/ACM 45th International Conference on Software Engineering}.\hskip 1em plus 0.5em minus 0.4em\relax IEEE, 2023, pp. 1443--1455.

\bibitem{jiang2019inferring}
J.~Jiang, L.~Ren, Y.~Xiong, and L.~Zhang, ``Inferring program transformations from singular examples via big code,'' in \emph{2019 34th IEEE/ACM International Conference on Automated Software Engineering (ASE)}.\hskip 1em plus 0.5em minus 0.4em\relax IEEE, 2019, pp. 255--266.

\bibitem{Xia2022LessTM}
C.~Xia and L.~Zhang, ``Less training, more repairing please: revisiting automated program repair via zero-shot learning,'' \emph{Proceedings of the 30th ACM Joint European Software Engineering Conference and Symposium on the Foundations of Software Engineering}, 2022.

\bibitem{xia2023plastic}
C.~S. Xia, Y.~Ding, and L.~Zhang, ``The plastic surgery hypothesis in the era of large language models,'' in \emph{2023 38th IEEE/ACM International Conference on Automated Software Engineering}, 2023, pp. 522--534.

\bibitem{zhang2023gamma}
Q.~Zhang, C.~Fang, T.~Zhang, B.~Yu, W.~Sun, and Z.~Chen, ``Gamma: Revisiting template-based automated program repair via mask prediction,'' in \emph{2023 38th IEEE/ACM International Conference on Automated Software Engineering}, 2023, pp. 535--547.

\bibitem{xiang2024far}
J.~Xiang, X.~Xu, F.~Kong, M.~Wu, H.~Zhang, and Y.~Zhang, ``How far can we go with practical function-level program repair?'' \emph{arXiv preprint arXiv:2404.12833}, 2024.

\bibitem{silva2023repairllama}
A.~Silva, S.~Fang, and M.~Monperrus, ``Repairllama: Efficient representations and fine-tuned adapters for program repair,'' \emph{arXiv preprint arXiv:2312.15698}, 2023.

\bibitem{xia2024automated}
C.~S. Xia and L.~Zhang, ``Automated program repair via conversation: Fixing 162 out of 337 bugs for \$0.42 each using chatgpt,'' in \emph{Proceedings of the 33rd ACM SIGSOFT International Symposium on Software Testing and Analysis}, 2024, pp. 819--831.

\bibitem{yin2024thinkrepair}
X.~Yin, C.~Ni, S.~Wang, Z.~Li, L.~Zeng, and X.~Yang, ``Thinkrepair: Self-directed automated program repair,'' in \emph{Proceedings of the 33rd ACM SIGSOFT International Symposium on Software Testing and Analysis}, 2024, pp. 1274--1286.

\bibitem{xia2023automated}
C.~S. Xia, Y.~Wei, and L.~Zhang, ``Automated program repair in the era of large pre-trained language models,'' in \emph{IEEE/ACM 45th International Conference on Software Engineering}, 2023, pp. 1482--1494.

\bibitem{jiang2023impact}
N.~Jiang, K.~Liu, T.~Lutellier, and L.~Tan, ``Impact of code language models on automated program repair,'' in \emph{2023 IEEE/ACM 45th International Conference on Software Engineering}.\hskip 1em plus 0.5em minus 0.4em\relax IEEE, 2023, pp. 1430--1442.

\bibitem{wei2022chain}
J.~Wei, X.~Wang, D.~Schuurmans, M.~Bosma, F.~Xia, E.~Chi, Q.~V. Le, D.~Zhou \emph{et~al.}, ``Chain-of-thought prompting elicits reasoning in large language models,'' \emph{Advances in neural information processing systems}, vol.~35, pp. 24\,824--24\,837, 2022.

\bibitem{yi2024survey}
Z.~Yi, J.~Ouyang, Y.~Liu, T.~Liao, Z.~Xu, and Y.~Shen, ``A survey on recent advances in llm-based multi-turn dialogue systems,'' \emph{arXiv preprint arXiv:2402.18013}, 2024.

\bibitem{xi2023rise}
Z.~Xi, W.~Chen, X.~Guo, W.~He, Y.~Ding, B.~Hong, M.~Zhang, J.~Wang, S.~Jin, E.~Zhou \emph{et~al.}, ``The rise and potential of large language model based agents: A survey,'' \emph{arXiv preprint arXiv:2309.07864}, 2023.

\bibitem{just2014defects4j}
R.~Just, D.~Jalali, and M.~D. Ernst, ``Defects4j: A database of existing faults to enable controlled testing studies for java programs,'' in \emph{Proceedings of the 2014 international symposium on software testing and analysis}, 2014, pp. 437--440.

\bibitem{lin2017quixbugs}
D.~Lin, J.~Koppel, A.~Chen, and A.~Solar-Lezama, ``Quixbugs: A multi-lingual program repair benchmark set based on the quixey challenge,'' in \emph{Proceedings Companion of the 2017 ACM SIGPLAN international conference on systems, programming, languages, and applications: software for humanity}, 2017, pp. 55--56.

\bibitem{zhang2023critical}
Q.~Zhang, T.~Zhang, J.~Zhai, C.~Fang, B.~Yu, W.~Sun, and Z.~Chen, ``A critical review of large language model on software engineering: An example from chatgpt and automated program repair,'' \emph{arXiv preprint arXiv:2310.08879}, 2023.

\bibitem{lopez2024inter}
J.~A.~H. L{\'o}pez, B.~Chen, T.~Sharma, and D.~Varr{\'o}, ``On inter-dataset code duplication and data leakage in large language models,'' \emph{arXiv preprint arXiv:2401.07930}, 2024.

\bibitem{sallou2024breaking}
J.~Sallou, T.~Durieux, and A.~Panichella, ``Breaking the silence: the threats of using llms in software engineering,'' in \emph{Proceedings of the 2024 ACM/IEEE 44th International Conference on Software Engineering: New Ideas and Emerging Results}, 2024, pp. 102--106.

\bibitem{xu2024benchmarking}
R.~Xu, Z.~Wang, R.-Z. Fan, and P.~Liu, ``Benchmarking benchmark leakage in large language models,'' \emph{arXiv preprint arXiv:2404.18824}, 2024.

\bibitem{homepage}
{Homepage}, 2024, available at: \url{https://zenodo.org/records/13901271}.

\bibitem{10.1145/3663529.3663815}
Y.~Wu, Z.~Li, J.~M. Zhang, and Y.~Liu, ``Condefects: A complementary dataset to address the data leakage concern for llm-based fault localization and program repair,'' in \emph{Companion Proceedings of the 32nd ACM International Conference on the Foundations of Software Engineering}, 2024, p. 642–646.

\bibitem{chen2021evaluating}
M.~Chen, J.~Tworek, H.~Jun, Q.~Yuan, H.~P. D.~O. Pinto, J.~Kaplan, H.~Edwards, Y.~Burda, N.~Joseph, G.~Brockman \emph{et~al.}, ``Evaluating large language models trained on code,'' \emph{arXiv preprint arXiv:2107.03374}, 2021.

\bibitem{li2024evocodebench}
J.~Li, G.~Li, X.~Zhang, Y.~Dong, and Z.~Jin, ``Evocodebench: An evolving code generation benchmark aligned with real-world code repositories,'' \emph{arXiv preprint arXiv:2404.00599}, 2024.

\bibitem{xia2024top}
C.~S. Xia, Y.~Deng, and L.~Zhang, ``Top leaderboard ranking= top coding proficiency, always? evoeval: Evolving coding benchmarks via llm,'' \emph{arXiv e-prints}, pp. arXiv--2403, 2024.

\bibitem{liu2024your}
J.~Liu, C.~S. Xia, Y.~Wang, and L.~Zhang, ``Is your code generated by chatgpt really correct? rigorous evaluation of large language models for code generation,'' \emph{Advances in Neural Information Processing Systems}, vol.~36, 2024.

\bibitem{EvalPlusLeaderBoard}
``Evalplus leaderboard,'' \url{https://evalplus.github.io/leaderboard.html}, 2024, accessed: 2024-6-05.

\bibitem{yu2024wavecoder}
Z.~Yu, X.~Zhang, N.~Shang, Y.~Huang, C.~Xu, Y.~Zhao, W.~Hu, and Q.~Yin, ``Wavecoder: Widespread and versatile enhancement for code large language models by instruction tuning,'' in \emph{Proceedings of the 62nd Annual Meeting of the Association for Computational Linguistics (Volume 1: Long Papers)}, 2024, pp. 5140--5153.

\bibitem{guo2024deepseek}
D.~Guo, Q.~Zhu, D.~Yang, Z.~Xie, K.~Dong, W.~Zhang, G.~Chen, X.~Bi, Y.~Wu, Y.~Li \emph{et~al.}, ``Deepseek-coder: When the large language model meets programming--the rise of code intelligence,'' \emph{arXiv preprint arXiv:2401.14196}, 2024.

\bibitem{zheng2024opencodeinterpreter}
T.~Zheng, G.~Zhang, T.~Shen, X.~Liu, B.~Y. Lin, J.~Fu, W.~Chen, and X.~Yue, ``Opencodeinterpreter: Integrating code generation with execution and refinement,'' \emph{arXiv preprint arXiv:2402.14658}, 2024.

\bibitem{wei2024magicoder}
Y.~Wei, Z.~Wang, J.~Liu, Y.~Ding, and L.~Zhang, ``Magicoder: Empowering code generation with oss-instruct,'' in \emph{Forty-first International Conference on Machine Learning}, 2024.

\bibitem{Artigenz-Coder}
``Artigenz-coder-ds,'' \url{https://huggingface.co/Artigenz/Artigenz-Coder-DS-6.7B}, 2024, accessed: 2024-6-05.

\bibitem{qwen}
J.~Bai, S.~Bai, Y.~Chu, Z.~Cui, K.~Dang, X.~Deng, Y.~Fan, W.~Ge, Y.~Han, F.~Huang \emph{et~al.}, ``Qwen technical report,'' \emph{arXiv preprint arXiv:2309.16609}, 2023.

\bibitem{Tunstall_The_Alignment_Handbook}
\BIBentryALTinterwordspacing
L.~Tunstall, E.~Beeching, N.~Lambert, N.~Rajani, S.~Huang, K.~Rasul, A.~Bartolome, A.~M.~Rush, and T.~Wolf, ``{The Alignment Handbook}.'' [Online]. Available: \url{https://github.com/huggingface/alignment-handbook}
\BIBentrySTDinterwordspacing

\bibitem{SpeechLess-CodeLlama-v2.0}
``Speechless-codellama-v2.0,'' \url{https://huggingface.co/uukuguy/speechless-codellama-34b-v2.0}, 2024, accessed: 2024-6-22.

\bibitem{cheers2019spplagiarise}
H.~Cheers, Y.~Lin, and S.~P. Smith, ``Spplagiarise: A tool for generating simulated semantics-preserving plagiarism of java source code,'' in \emph{2019 IEEE 10th International conference on software engineering and service science (ICSESS)}.\hskip 1em plus 0.5em minus 0.4em\relax IEEE, 2019, pp. 617--622.

\bibitem{tian2023code}
Z.~Tian, J.~Chen, and Z.~Jin, ``Code difference guided adversarial example generation for deep code models,'' in \emph{2023 38th IEEE/ACM International Conference on Automated Software Engineering (ASE)}.\hskip 1em plus 0.5em minus 0.4em\relax IEEE, 2023, pp. 850--862.

\bibitem{JDTAST}
``{Eclipse JDT Core},'' \url{https://www.eclipse.org/jdt/core/}, Eclipse Foundation, 2024, accessed: 2024-06.

\bibitem{HuggingFace}
``Hugging face,'' \url{https://huggingface.co}, 2024, accessed: 2024-6-04.

\end{thebibliography}

\end{document}